\documentclass[twocolumn,aps,prb,superscriptaddress,a4paper,showpacs,showkeys]{revtex4-1}

\usepackage{graphicx}% Include figure files
\usepackage{bm,amssymb,amsmath,dcolumn}
\usepackage{url,cancel}
\usepackage{type1cm}% scalable Adobe fonts only
\usepackage[dvips]{color}
\usepackage[normalem]{ulem}

\newcommand{\mm}[1]{\mbox{$#1$}}
\newcommand{\dstd}{\mathrm{d}}
\newcommand{\istd}{\mathrm{i}}

\newcommand{\dva}{$L_{2}$~edge}
\newcommand{\dvatri}{$L_{2,3}$~edge}
\newcommand{\tri}{$L_{3}$~edge}
\newcommand{\xass}{x-ray absorption spectra}

\newcommand{\lr}{Lorentzian}
\newcommand{\ef}{$E_{F}$}

\newcommand{\ea}{{\it et al.}}

\begin{document}

\title{Finite lifetime broadening of calculated x-ray absorption
  spectra: possible artefacts close to the edge}

\author{Ond\v{r}ej \v{S}ipr}
\email{sipr@fzu.cz}
\homepage{http://www.fzu.cz/~sipr} 
\affiliation{Institute of Physics, Czech Academy of Sciences,
  Cukrovarnick\'{a}~10, CZ-162~53~Prague, Czech Republic }

\author{Ji\v{r}\'{\i} Vack\'{a}\v{r}}
\affiliation{Institute of Physics, Czech Academy of Sciences,
  Na~Slovance~2, CZ-182~21~Prague, Czech Republic}

\author{J\'{a}n Min\'{a}r}
\affiliation{University of West Bohemia, Univerzitn\'{\i} 8,
  CZ-306~14~Pilsen, Czech Republic}

\date{\today}

%%%%%%%%%%%%%%%%%%%%%%%%%%%%%%%%%%%%%%%%%%%%%%%%%%%%%%%%%%%%%%%

\begin{abstract}
X-ray absorption spectra calculated within an effective one-electron
approach have to be broadened to account for the finite lifetime of the
core hole.  For Green's function based methods this can be achieved
either by adding a small imaginary part to the energy or by
convoluting the spectra on the real axis with a Lorentzian.  We
demonstrate on the case of Fe $K$ and \dvatri\ spectra that these
procedures lead to identical results only for energies higher than few
core level widths above the absorption edge. For energies close to the
edge, spurious spectral features may appear if too much weight is put
on broadening via the imaginary energy component.  Special care should
be taken for dichroic spectra at edges which comprise several
exchange-split core levels, such as the $L_{3}$ edge of 3$d$
transition metals.
\end{abstract}


%%%%%%%%%%%%%%%%%%%%%%%%%%%%%%%%%%%%%%%%%%%%%%%%%%%%%%%%%%%%%%%

\maketitle         

%%%%%%%%%%%%%%%%%%%%%%%%%%%%%%%%%%%%%%%%%%%%%%%%%%%%%%%%%%%%%%%

\section{Introduction}

\label{sec-intro}

Generally, experimental x-ray absorption spectra (XAS) contain fewer
structures and display broader features than theoretical spectra.
This is because the finite lifetime of the core hole is usually
neglected in the calculations.  To facilitate proper comparison
between theory and experiment, the calculated spectrum is modified so
that the finite core hole lifetime is accounted for.  A convenient way
to achieve this is to convolute the raw spectrum a posteriori with a
Lorentzian.\cite{Messiah}  This is a well-established procedure.  Its
drawback is that one has to perform the calculations on sometimes much
finer energy mesh than actually needed: the raw spectrum contains many
fine and sharp features that will be smeared out eventually but which,
nevertheless, have to be included in the calculated spectrum before
the final broadening is applied.

For Green's function based or multiple-scattering methods, there is
another --- computationally more efficient --- way to account for the
finite core hole lifetime, namely, adding a small imaginary part to
the energy.\cite{Messiah,VGD+82,BAB96,NBD03,SGW+06}  This will result in
smoother spectra from the beginning, meaning that one can use a
coarser energy mesh (cf.\ an instructive demonstration presented
recently by Taranukhina \ea).\cite{TNK+18} 
Another technical advantage of this approach from a
computational viewpoint is that when working in a reciprocal space,
employing complex energies may significantly reduce the number of
$\bm{k}$-points needed for an accurate Brillouin zone integration.
The option to use complex energies is available in several codes
designed for XAS calculations, among others, in {\sc
  fdmnes},\cite{fdmnes-code,BJ+09} {\sc feff},\cite{feff-code,RKP+09}
{\sc MsSpec},\cite{msspec-code,SNG+11} {\sc
  mxan},\cite{mxan-code,BDN+03} or {\sc
  sprkkr}.\cite{sprkkr-code,EKM11} 
Often one can combine both approaches by first calculating the
spectrum using an imaginary energy component to achieve basic
reduction of the computer workload and then by convoluting it with a
Lorentzian to achieve the best possible agreement with experiment.

The problem with calculating \xass\ for energies with an added
imaginary component is that this procedure is formally equivalent to a
convolution with a Lorentzian only if there is no cut-off of the
spectra below the Fermi level~\ef, i.e., in the limit \mm{E_{F}
  \rightarrow -\infty}.\cite{VGD+82,BAB96}  Usually this circumstance is
tacitly ignored because the influence of the cut-off is negligible
sufficiently above the edge.  Specifically, this applies to the whole
extended x-ray absorption fine structure (EXAFS) region.  However, the
question remains whether and under what circumstances the employment
of an imaginary energy component might lead to undesirable artefacts
at the very edge. Brouder \ea\cite{BAB96} derived an equation linking XAS
calculated for complex energies to XAS obtained by convolution of
spectra on the real axis which takes into account the influence of the
cut-off below \ef.  Their equation contains a correction factor which
could be in principle calculated but which is normally ignored.  An
analysis of how serious this neglect might be has not been presented
so far.

The finite core hole lifetime is not the only factor that contributes
to the broadening of spectra. Other effects to consider are, e.g.,
finite lifetime of the excited photoelectron or thermal
vibrations. The broadening caused by the finite core hole lifetime is,
nevertheless, the dominant broadening process close to the edge and
knowing the limitations of procedures used to account for it is
desirable.

Our aim is to assess whether employment of an imaginary energy
component to calculate broadened \xass\ can introduce significant
artefacts in comparison with broadening by a convolution of raw
spectra calculated on the real axis.  To cover a range of
circumstances, we investigate Fe $K$ edge and Fe \dvatri\ XAS and
x-ray magnetic circular dichroism (XMCD). We will show that if the
dominant mechanism of the broadening is adding an imaginary part to
the energy, spurious spectral features may appear close to the edge.
Especially this is the case of dichroic spectra at edges which
comprise several exchange-split core levels of small natural widths,
as it is the case, e.g., of $L_{3}$ edges of 3$d$ transition metals.

%%%%%%%%%%%%%%%%%%%%%%%%%%%%%%%%%%%%%%%%%%%%%%%%%%%%%%%%%%%%%%%
%%%%%%%%%%%%%%%%%%%%%%%%%%%%%%%%%%%%%%%%%%%%%%%%%%%%%%%%%%%%%%%

\section{Methods}

Fe $K$ edge and Fe \dvatri\ XAS and XMCD spectra were calculated using
an ab-initio fully-relativistic multiple-scattering Green's function
method, as implemented in the {\sc sprkkr}
code.\cite{sprkkr-code,EKM11}
We are dealing with crystals, so the
calculations were done in the reciprocal space.  The $\bm{k}$-space
integrals were carried out using 36000 points in the full Brillouin
zone.  Multipole expansion of the Green's function was cut at
$\ell_{\text{max}}$=2.  We checked that these values are sufficient.
The influence of the core hole on the potential was ignored, which is
justified for our purpose; quantitative estimate of the core hole
effect can be found, e.g., in Zeller\cite{Zel88} for the Fe $K$ edge
and in \v{S}ipr \ea\cite{SMS+11} for the Fe \dvatri.

Use of fully relativistic formalism means that the core levels
associated with absorption edges are non-degenerate, separated by
exchange splitting.  For the $K$ edge, core levels characterized by
relativistic quantum numbers $\kappa=-1$, $\mu=\pm1/2$ are split by
0.005~eV.  For the \dva, core levels characterized by $\kappa=1$,
$\mu=\pm1/2$ are split by 0.3~eV.  For the \tri, the four levels
characterized by $\kappa=-2$, $\mu=\pm1/2,\pm3/2$ are also split by
about 0.3~eV from each other, spanning the total range of 1~eV.  We
will see that this exchange splitting of core levels contributes to
possible emergence of artefacts close to the edge if the spectra are
broadened by employing complex energies.

As noted in the Introduction, there are two ways to simulate the
effect of the finite core hole lifetime on \xass.  First, it is the
convolution of the raw spectrum calculated for real photoelectron
energies by a Lorentzian $L(E)$.  If the full width at half maximum
(FWHM) of the \lr\ is $\Gamma$, it can be written as
\begin{equation}
  L(E) \: = \: \frac{1}{\pi} \, 
  \frac{\Gamma/2}{E^{2} + (\Gamma/2)^{2}}
\quad .
\label{eq-lor}
\end{equation}
Starting with a raw x-ray
absorption cross-section  $\sigma_{\text{raw}}(E)$ which ignores the 
finite core lifetime effects, one makes a convolution
\begin{equation}
  \sigma(E) \: = \:
  \int_{E_{F}}^{\infty} \! \dstd E' \:
  \sigma_{\text{raw}}(E') \:  L(E-E')
\label{eq-conv}
\end{equation}
to obtain the cross-section $\sigma(E)$ where the influence of the finite
core hole lifetime has been included.

It can be shown that if the cut-off at \ef\ is ignored in
Eq.~(\ref{eq-conv}), the effect of the finite core hole lifetime can
be equivalently accounted for by evaluating the x-ray absorption
cross-section for energies with added imaginary component
$\Gamma/2$.\cite{Messiah,BAB96,NBD03,SGW+06}  In other words, 
\begin{equation}
  \int_{-\infty}^{\infty} \! \dstd E' \:
  \sigma_{\text{raw}}(E') \:  L(E-E')
  \: = \:
  \sigma_{\text{raw}}(E+\istd\Gamma/2)
  \quad .
  \label{eq-equiv}
\end{equation}
The x-ray absorption cross-section with the influence of finite core
lifetime included is thus taken as
\begin{equation}
  \sigma(E) \: \approx \: \sigma_{\text{raw}}(E+\istd\Gamma/2)
  \quad .
  \label{eq-imag}
\end{equation}

We want to test to what degree one can use Eq.~(\ref{eq-imag}) instead
of Eq.~(\ref{eq-conv}), saving thus computer resources and increasing
the numerical stability of the results.  For this purpose we
distribute the required core level broadening between the procedures
described by Eq.~(\ref{eq-conv}) and Eq.~(\ref{eq-imag}), with various
weights.  This can be done consistently because a convolution of two
Lorentzians with FWHM's of $\Gamma_{1}$ and $\Gamma_{2}$ is again a
Lorentzian, with FWHM equal to $\Gamma_{1}+\Gamma_{2}$.  So if we want
to simulate a total core hole broadening $\Gamma_{\text{core}}$, we
employ first Eq.~(\ref{eq-imag}) with an imaginary energy part
$\text{Im}E$ and then convolute the spectrum with a Lorentzian of
width $\Gamma$, requiring that
\begin{equation}
  2 \, \text{Im}E \: + \: \Gamma \: = \: \Gamma_{\text{core}}
\quad .
\label{eq-sum}
\end{equation}

\begin{table}
\caption{Parameters (in eV) used for broadening the Fe $K$ edge
  spectra by including the imaginary part $\text{Im}E$ to the energy
  and by subsequent convolution of the calculated spectra with a
  Lorentzian of the width $\Gamma_{K}$. Values are chosen so that the
  total (combined) broadening determined by Eq.~(\protect\ref{eq-sum})
  corresponds to~$\Gamma_{\text{core}}=1.19$~eV.  }
\begin{tabular}{ccc}
\hline \hline 
 \mbox{$\text{Im}E$} & 
 \mbox{$\Gamma_{K}$}  & 
 \mbox{$2\text{Im}E+\Gamma_{K}$} \\
\hline
  0.014   &  1.163  &  1.190  \\
  0.272   &  0.646  &  1.190  \\
  0.408   &  0.374  &  1.190  \\
  0.594   &  0.002  &  1.190  \\
\hline \hline 
\end{tabular}
\label{tab-kwidth}
\end{table}

In our case we take $\Gamma_{\text{core}}=1.19$~eV for the Fe $K$
edge, $\Gamma_{\text{core}}=1.14$~eV for the Fe \dva\ and
$\Gamma_{\text{core}}=0.41$~eV for the Fe \tri.\cite{CP01}  The
values of $\text{Im}E$ and $\Gamma$ used for studying the Fe $K$ edge
are shown in Tab.~\ref{tab-kwidth}, the values used for studying the
Fe \dvatri\ are shown in Tab.~\ref{tab-l23width}.  The rationale for
selecting these particular values is that in this way we include both
extreme situations when the broadening is incorporated either solely
via the Lorentzian convolution or solely via the imaginary energy
component, together with two intermediate situations.  Note that for
the \dvatri\ spectra the convolution is done separately for the
\dva\ and \tri\ and only then both spectra are merged.

\begin{table}
\caption{As Tab.~\protect\ref{tab-kwidth} but for the $L_{2,3}$
  edge. Values are chosen so that the total broadening determined by
  Eq.~(\protect\ref{eq-sum}) corresponds to
  $\Gamma_{\text{core}}=1.14$~eV for the \dva\ and to
  $\Gamma_{\text{core}}=0.41$~eV for the \tri.  }
\begin{tabular}{ccccc}
\hline \hline \\
 \mbox{$\text{Im}E$} & %
 \mbox{$\Gamma_{L_{2}}$} & %
 \mbox{$\Gamma_{L_{3}}$} &
 \mbox{$2\text{Im}E+\Gamma_{L_{2}}$} & %
 \mbox{$2\text{Im}E+\Gamma_{L_{3}}$} \\
\hline
  0.014   &  1.113  &  0.383 &  1.140  &  0.410  \\
  0.068   &  1.004  &  0.274 &  1.140  &  0.410  \\
  0.136   &  0.868  &  0.138 &  1.140  &  0.410  \\
  0.204   &  0.732  &  0.002 &  1.140  &  0.410  \\
\hline \hline 
\end{tabular}
\label{tab-l23width}
\end{table}

%%%%%%%%%%%%%%%%%%%%%%%%%%%%%%%%%%%%%%%%%%%%%%%%%%%%%%%%%%%%%%%
%%%%%%%%%%%%%%%%%%%%%%%%%%%%%%%%%%%%%%%%%%%%%%%%%%%%%%%%%%%%%%%

\section{Results: Fe $K$ edge and $L_{2,3}$ edge XAS and XMCD}

\label{sec-results}

%%---%%---%%---%%---%%---%%---%%---%%---%%---%%---%%---%%---%%

\subsection{Total spectra (unresolved in $\kappa$ or $\mu$)}

\label{sec-xasxmcd}

% Figure planned for 1 column
\begin{figure}
 \includegraphics[viewport=0.0cm 0.0cm 9.0cm
   13.5cm,width=85mm]{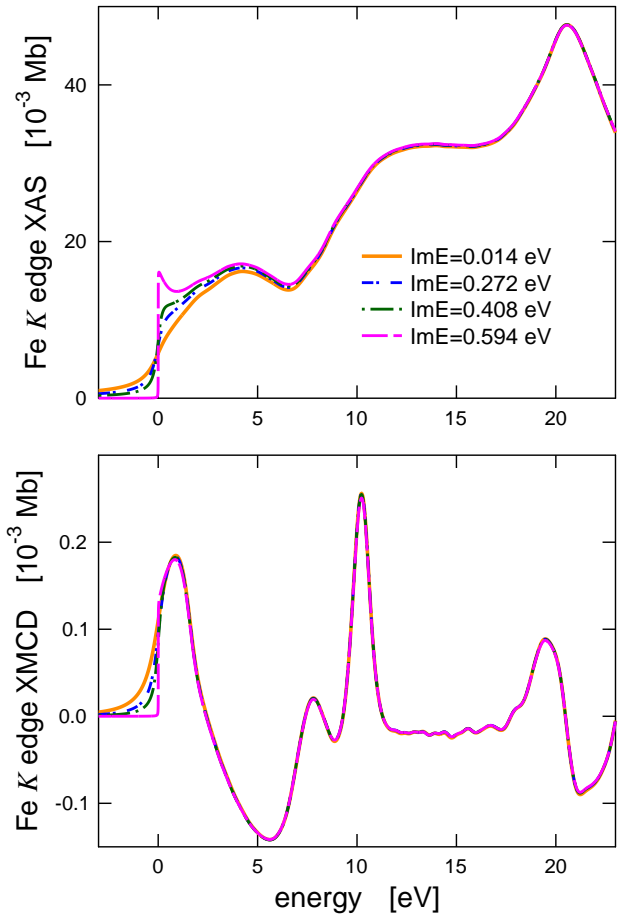}
 \caption{Fe $K$ edge XAS and XMCD calculated for different imaginary
  energies and convoluted subsequently with Lorentzians chosen so that
  the total spectral broadening $2\text{Im}E+\Gamma_{K}$ remains
  constant.} 
\label{fig-kedge}
\end{figure}

\begin{figure}
\includegraphics[viewport=0.0cm 0.0cm 9.0cm
  13.5cm,width=85mm]{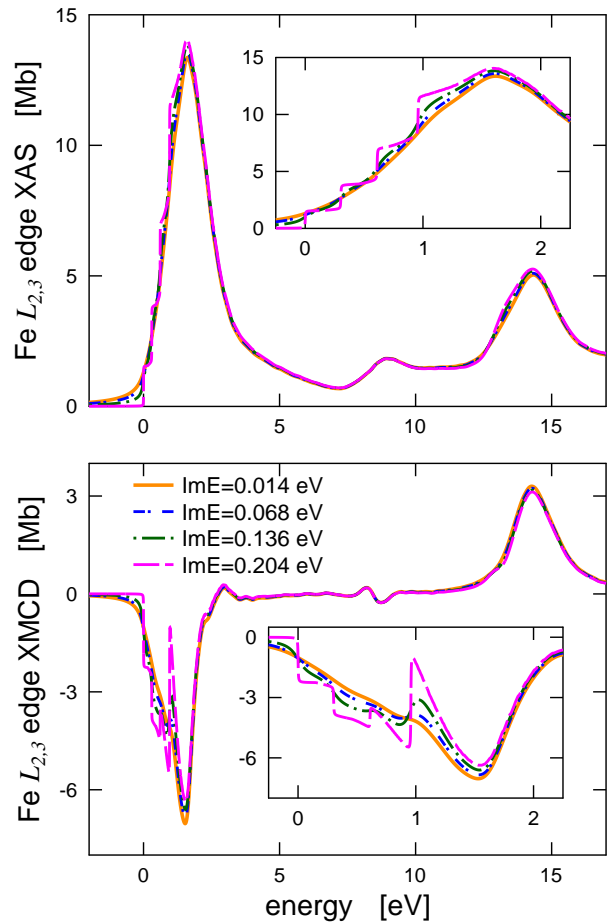} 
\caption{As Fig.~\protect\ref{fig-kedge} but for the
  $L_{2,3}$ edge.  The insets show detailed views on the \tri\ region.}
\label{fig-l23edge}
\end{figure}

We start by inspecting how shifting the weight of the broadening from
a Lorentzian convolution to an imaginary energy component affects the
calculated XAS and XMCD.  This is done in Fig.~\ref{fig-kedge} for the
Fe $K$ edge and in Fig.~\ref{fig-l23edge} for the Fe \dvatri.  We
focus solely on the theoretical spectra; a good agreement of theory
with experiment was demonstrated earlier.\cite{SE05,SMS+11}

% Figure planned for 1 column

One can see that for energies higher than about \mm{5 \,
  \Gamma_{\text{core}}}\ above the edge there is practically no
difference in the spectra, no matter which broadening procedure has
been applied.  At the very edge, however, there are differences. They
stem from the fact that if too much weight is put on broadening by
means of the imaginary energy component, there is a sharp cut-off of
the spectra at \ef, resulting in too sharp features at the edge.  If a
sufficient amount of broadening is done via Lorentzian convolution,
this cut-off is smeared out.

The situation is especially instructive for the Fe \tri\ XMCD peak.
Here a well-distinguished but in fact spurious fine structure appears
on its low-energy side unless most of the broadening is done by means
of Lorentzian convolution.  The situation is much less dramatic for
the corresponding XAS peak.  This is because of the way the XMCD peak
is generated: it is a sum of four (\tri) or two (\dva) contributions
which may have opposite signs and which have their edges at slightly
different energies due to the relativistic exchange splitting of the
core levels.

%%---%%---%%---%%---%%---%%---%%---%%---%%---%%---%%---%%---%%

\subsection{($\kappa$,$\mu$)-resolved spectra}

\label{sec-kappamu}

% Figure planned for 1 column
\begin{figure}
  \includegraphics[viewport=0.0cm 0.0cm 9.0cm
    13.5cm,width=85mm]{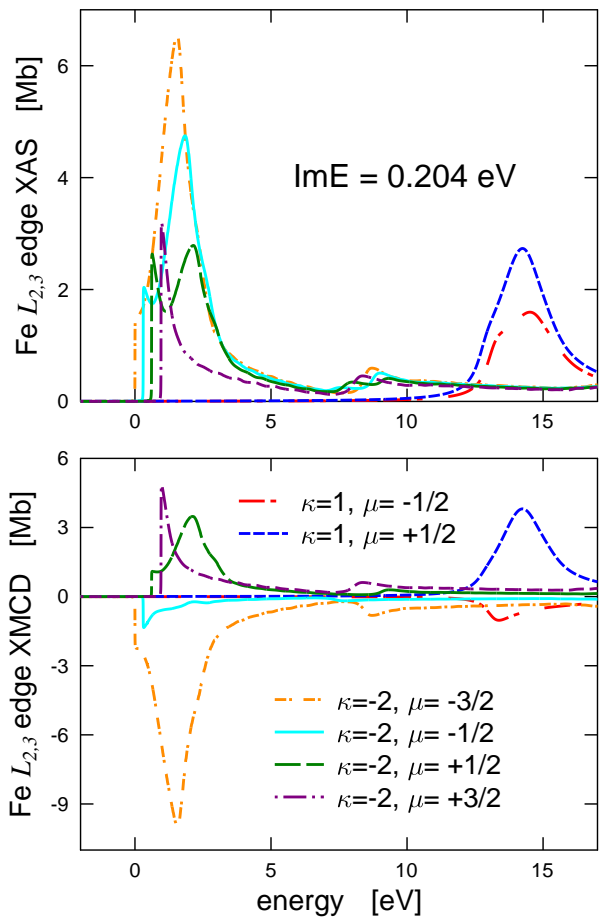} 
\caption{($\kappa$,$\mu$)-decomposed Fe $L_{2,3}$ edge XAS and XMCD
  calculated for imaginary energy 0.204~eV and convoluted subsequently
  with Lorentzians of $\Gamma_{L_{2}}=0.732$~eV (for the \dva) and
  $\Gamma_{L_{3}}=0.002$~eV (for the \tri).}
\label{fig-decomp-020}
\end{figure}

% Figure planned for 1 column
\begin{figure}
\includegraphics[viewport=0.0cm 0.0cm 9.0cm
  13.5cm,width=85mm]{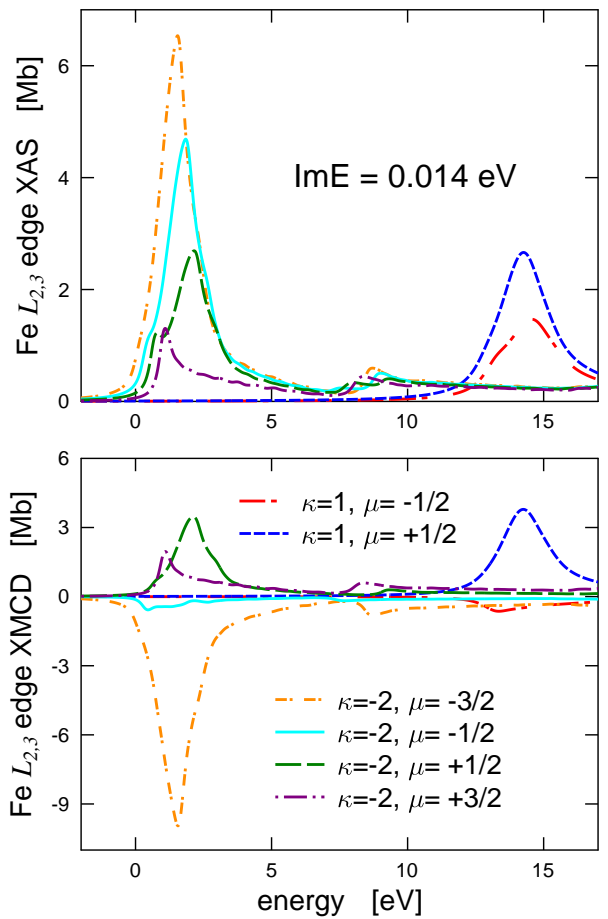}
\caption{As Fig.~\protect\ref{fig-decomp-020} but for for imaginary
  energy 0.014~eV and Lorentzian widths $\Gamma_{L_{2}}=1.113$~eV and
  $\Gamma_{L_{3}}=0.383$~eV for the $L_{2}$ and $L_{3}$ edges,
  respectively.} 
\label{fig-decomp-001}
\end{figure}

An insight can be obtained by looking at individual
($\kappa$,$\mu$)-components contributing to the Fe \dvatri\ XAS and
XMCD for the two extreme cases, with $\text{Im}E=0.204$~eV (nearly all
broadening done via the imaginary energy component) and with
$\text{Im}E=0.014$~eV (nearly all broadening done via a convolution
with a Lorentzian --- cf.\ Tab.~\ref{tab-l23width}). If one looks at
the spectra obtained using $\text{Im}E=0.204$~eV
(Fig.~\ref{fig-decomp-020}), one can see that the individual
components indeed exhibit sharp edges or onsets at different energies.
If all components have the same sign, as is the case of XAS, the
resulting spectrum is ``rugged'' but in general not so much different
from the spectra obtained for smaller $\text{Im}E$ (see the upper
panel in Fig.~\ref{fig-l23edge}).  However, if the individual
($\kappa$,$\mu$) components differ in signs as in the case of XMCD,
their sum may give rise to a spectrum which is significantly different
from the spectrum obtained for smaller $\text{Im}E$ (the lower panel
in Fig.~\ref{fig-l23edge}).  Technically, this difference can be
understood by comparing Fig.~\ref{fig-decomp-020} with
Fig.~\ref{fig-decomp-001}, which is its analog but with a smaller
imaginary energy part $\text{Im}E=0.014$~eV.  One case see that if the
individual ($\kappa$,$\mu$)-components have been smoothed {\em before}
the summation, the resulting spectrum is smooth as well, without the
quasi-oscillation at about 1~eV which appears in the XMCD spectrum for
larger $\text{Im}E$'s (cf.\ Fig.~\ref{fig-l23edge}).

%%%%%%%%%%%%%%%%%%%%%%%%%%%%%%%%%%%%%%%%%%%%%%%%%%%%%%%%%%%%%%%%%%%%%%%
%%%%%%%%%%%%%%%%%%%%%%%%%%%%%%%%%%%%%%%%%%%%%%%%%%%%%%%%%%%%%%%%%%%%%%%

\section{Discussion}

Our aim was to check whether employment of complex energies for
calculating broadened XAS and XMCD spectra can introduce significant
distortions in comparison with what is obtained by convoluting the
spectra calculated on the real axis.  Our results indicate that
simulating the finite core hole lifetime by means of an imaginary
energy component and by means of convoluting the raw spectra with a
Lorentzian is equivalent only for energies higher than few core level
FWHM's above the absorption edge.  If too much weight is put on
broadening via an imaginary energy component, spurious spectral
features may appear close to the edge, especially for the dichroic
spectra.

When contemplating practical implications of this, one should realize
that there are other sources of spectral broadening that we did not
consider.  In particular the finite lifetime of the
photoelectron\cite{MJW82} and also atomic
vibrations.\cite{BP+76,FRW+99,SVK+16} 
The influence of these effects will be, nevertheless, small at the
very edge, which is the region where there is the largest likelihood
that the ansatz Eq.~(\ref{eq-imag}) will be inappropriate.  Another
factor important for comparison with experiment is the instrumental
broadening.  This is usually accounted for by a Gaussian smearing.
Typical values for the width of the Gaussian are 0.8~eV for the Fe $K$
edge and 0.2~eV for the Fe \dvatri.  Applying the instrumental
broadening would thus remove most of the significant differences
between individual spectra shown in Fig.~\ref{fig-kedge} and also
between the XAS spectra shown in the upper panel of
Fig.~\ref{fig-l23edge}.  However, the spurious fine structure
appearing for large $\text{Im}E$'s at the low-energy side of the Fe
\tri\ XMCD peak would remain.

Typical values of the imaginary energy component used for XAS/XMCD
calculations are \mbox{$\text{Im}E \sim 0.1$--0.2}~eV.  It follows
from our results that while this is appropriate for most situations,
problems might occur for edges where the core hole lifetime broadening
is small --- such as the $L_{3}$ edges of 3$d$ transition metals.
Special care should be taken for XMCD spectra: it is still reasonable
to perform the calculations for complex energies to reduce the
computing workload but $\text{Im}E$ should be smaller than usually.
To give a specific recommendation, we suggest that $\text{Im}E$ should
be about one tenth of the tabulated FWHM value, i.e.,
0.1\mbox{$\times\Gamma_{\text{core}}$}.  An energy mesh dense enough
to account for all fine features that might be present in the spectrum
should have a step of $\text{Im}E$/2.

The severity of the effect investigated here increases if the applied
core hole lifetime broadening $\Gamma_{\text{core}}$ decreases.  For
example, if we had used FWHM's recommended by an older compilation of
Al Shamma \ea\cite{SAB+92} which are about 50~\% less than the values
recommended by the newer compilation of Campbell and Papp,\cite{CP01}
the spurious fine structure at the Fe \tri\ XMCD peak would be even
more pronounced. On the other hand, the fact that core hole lifetime
widths are usually not known very accurately means that the effect
explored here may be overlooked: if redundant structures appear close
to the edge, one might be tempted to apply mechanically additional
broadening, without considering that one may be actually dealing with
an artefact caused by the mechanism analyzed here.

Finally, the fact that both ways of dealing with core hole lifetime
broadening are equivalent sufficiently high above the edge justifies
formally the use of exponential damping in the EXAFS
region.\cite{LCE+81}
Namely, it can be shown that if the free-electron
Green's function is evaluated for a complex energy
\[  E \: + \: \istd \, \Gamma/2   \quad , \]
it gives rise to exponential damping of the photoelectron probability
with the mean free path $\lambda$,\cite{MJW82,NBD03,SGW+06} 
\[  \lambda \: = \: \frac{\hbar}{\Gamma} \, \sqrt{ \frac{2E}{m} }
  \quad . \] This leads straightforwardly to the $\exp(-R/\lambda)$
  factor used in the EXAFS formula.\cite{LCE+81,NBD03}  One only has
  to take care whether the mean free path $\lambda$ is related to the
  photoelectron probability as it is the case here or whether it is
  related to the amplitude [then the proper factor is
    $\exp(-2R/\lambda)$].

%%%%%%%%%%%%%%%%%%%%%%%%%%%%%%%%%%%%%%%%%%%%%%%%%%%%%%%%%%%%%%%%%%%%%%%
%%%%%%%%%%%%%%%%%%%%%%%%%%%%%%%%%%%%%%%%%%%%%%%%%%%%%%%%%%%%%%%%%%%%%%%

\section{Conclusions}

Well above the absorption edge, the two ways of incorporating the
finite core hole lifetime into calculation of x-ray absorption
spectra, namely, via adding an imaginary component to the energy and
via convoluting the raw spectrum with a Lorentzian, are equivalent.
However, this is not the case close to the edge.  Ignoring this can
lead to emergence of spurious spectral features.  Special care should
be taken for dichroic spectra at edges which comprise several
exchange-split core levels, as is the case of the $L_{3}$ edge of 3$d$
transition metals.

%%%%%%%%%%%%%%%%%%%%%%%%%%%%%%%%%%%%%%%%%%%%%%%%%%%%%%%%%%%%%%%

\begin{acknowledgements}
  This work was supported by the M\v{S}MT LD-COST~CZ project
  LD15097, by the GA~\v{C}R project 17-14840~S and by the CEDAMNF
  project CZ.02.1.01/0.0/0.0/15\_003/0000358.
\end{acknowledgements}

% Produces the bibliography via BibTeX.

\bibliography{liter-broadening}

\begin{thebibliography}{26}%
\makeatletter
\providecommand \@ifxundefined [1]{%
 \@ifx{#1\undefined}
}%
\providecommand \@ifnum [1]{%
 \ifnum #1\expandafter \@firstoftwo
 \else \expandafter \@secondoftwo
 \fi
}%
\providecommand \@ifx [1]{%
 \ifx #1\expandafter \@firstoftwo
 \else \expandafter \@secondoftwo
 \fi
}%
\providecommand \natexlab [1]{#1}%
\providecommand \enquote  [1]{``#1''}%
\providecommand \bibnamefont  [1]{#1}%
\providecommand \bibfnamefont [1]{#1}%
\providecommand \citenamefont [1]{#1}%
\providecommand \href@noop [0]{\@secondoftwo}%
\providecommand \href [0]{\begingroup \@sanitize@url \@href}%
\providecommand \@href[1]{\@@startlink{#1}\@@href}%
\providecommand \@@href[1]{\endgroup#1\@@endlink}%
\providecommand \@sanitize@url [0]{\catcode `\\12\catcode `\$12\catcode
  `\&12\catcode `\#12\catcode `\^12\catcode `\_12\catcode `\%12\relax}%
\providecommand \@@startlink[1]{}%
\providecommand \@@endlink[0]{}%
\providecommand \url  [0]{\begingroup\@sanitize@url \@url }%
\providecommand \@url [1]{\endgroup\@href {#1}{\urlprefix }}%
\providecommand \urlprefix  [0]{URL }%
\providecommand \Eprint [0]{\href }%
\providecommand \doibase [0]{http://dx.doi.org/}%
\providecommand \selectlanguage [0]{\@gobble}%
\providecommand \bibinfo  [0]{\@secondoftwo}%
\providecommand \bibfield  [0]{\@secondoftwo}%
\providecommand \translation [1]{[#1]}%
\providecommand \BibitemOpen [0]{}%
\providecommand \bibitemStop [0]{}%
\providecommand \bibitemNoStop [0]{.\EOS\space}%
\providecommand \EOS [0]{\spacefactor3000\relax}%
\providecommand \BibitemShut  [1]{\csname bibitem#1\endcsname}%
\let\auto@bib@innerbib\@empty
%</preamble>
\bibitem [{\citenamefont {Messiah}(1962)}]{Messiah}%
  \BibitemOpen
  \bibfield  {author} {\bibinfo {author} {\bibfnamefont {A.}~\bibnamefont
  {Messiah}},\ }\href@noop {} {\emph {\bibinfo {title} {Quantum Mechanics}}}\
  (\bibinfo  {publisher} {North-Holland},\ \bibinfo {address} {Amsterdam},\
  \bibinfo {year} {1962})\ \bibinfo {note} {vol.~2, p.~992}\BibitemShut
  {NoStop}%
\bibitem [{\citenamefont {Vedrinskii}\ \emph {et~al.}(1982)\citenamefont
  {Vedrinskii}, \citenamefont {Gegusin}, \citenamefont {Datsyuk}, \citenamefont
  {Novakovich},\ and\ \citenamefont {Kraizman}}]{VGD+82}%
  \BibitemOpen
  \bibfield  {author} {\bibinfo {author} {\bibfnamefont {R.~V.}\ \bibnamefont
  {Vedrinskii}}, \bibinfo {author} {\bibfnamefont {I.~I.}\ \bibnamefont
  {Gegusin}}, \bibinfo {author} {\bibfnamefont {V.~N.}\ \bibnamefont
  {Datsyuk}}, \bibinfo {author} {\bibfnamefont {A.~A.}\ \bibnamefont
  {Novakovich}}, \ and\ \bibinfo {author} {\bibfnamefont {V.~L.}\ \bibnamefont
  {Kraizman}},\ }\href {\doibase 10.1002/pssb.2221110202} {\bibfield  {journal}
  {\bibinfo  {journal} {phys. stat. sol. (b)}\ }\textbf {\bibinfo {volume}
  {111}},\ \bibinfo {pages} {433} (\bibinfo {year} {1982})}\BibitemShut
  {NoStop}%
\bibitem [{\citenamefont {Brouder}\ \emph {et~al.}(1996)\citenamefont
  {Brouder}, \citenamefont {Alouani},\ and\ \citenamefont {Bennemann}}]{BAB96}%
  \BibitemOpen
  \bibfield  {author} {\bibinfo {author} {\bibfnamefont {C.}~\bibnamefont
  {Brouder}}, \bibinfo {author} {\bibfnamefont {M.}~\bibnamefont {Alouani}}, \
  and\ \bibinfo {author} {\bibfnamefont {K.~H.}\ \bibnamefont {Bennemann}},\
  }\href@noop {} {\bibfield  {journal} {\bibinfo  {journal} {Phys. Rev. B}\
  }\textbf {\bibinfo {volume} {54}},\ \bibinfo {pages} {7334} (\bibinfo {year}
  {1996})}\BibitemShut {NoStop}%
\bibitem [{\citenamefont {Natoli}\ \emph {et~al.}(2003)\citenamefont {Natoli},
  \citenamefont {Benfatto}, \citenamefont {{Della Longa}},\ and\ \citenamefont
  {Hatada}}]{NBD03}%
  \BibitemOpen
  \bibfield  {author} {\bibinfo {author} {\bibfnamefont {C.~R.}\ \bibnamefont
  {Natoli}}, \bibinfo {author} {\bibfnamefont {M.}~\bibnamefont {Benfatto}},
  \bibinfo {author} {\bibfnamefont {S.}~\bibnamefont {{Della Longa}}}, \ and\
  \bibinfo {author} {\bibfnamefont {K.}~\bibnamefont {Hatada}},\ }\href@noop {}
  {\bibfield  {journal} {\bibinfo  {journal} {J.\ Synchr.\ Rad.}\ }\textbf
  {\bibinfo {volume} {10}},\ \bibinfo {pages} {26} (\bibinfo {year}
  {2003})}\BibitemShut {NoStop}%
\bibitem [{\citenamefont {S\'{e}billeau}\ \emph {et~al.}(2006)\citenamefont
  {S\'{e}billeau}, \citenamefont {Gunnella}, \citenamefont {Wu}, \citenamefont
  {Matteo},\ and\ \citenamefont {Natoli}}]{SGW+06}%
  \BibitemOpen
  \bibfield  {author} {\bibinfo {author} {\bibfnamefont {D.}~\bibnamefont
  {S\'{e}billeau}}, \bibinfo {author} {\bibfnamefont {R.}~\bibnamefont
  {Gunnella}}, \bibinfo {author} {\bibfnamefont {Z.-Y.}\ \bibnamefont {Wu}},
  \bibinfo {author} {\bibfnamefont {S.~D.}\ \bibnamefont {Matteo}}, \ and\
  \bibinfo {author} {\bibfnamefont {C.~R.}\ \bibnamefont {Natoli}},\ }\href
  {http://stacks.iop.org/0953-8984/18/i=9/a=R01} {\bibfield  {journal}
  {\bibinfo  {journal} {J. Phys.: Condens. Matter}\ }\textbf {\bibinfo {volume}
  {18}},\ \bibinfo {pages} {R175} (\bibinfo {year} {2006})}\BibitemShut
  {NoStop}%
\bibitem [{\citenamefont {Taranukhina}\ \emph {et~al.}(2018)\citenamefont
  {Taranukhina}, \citenamefont {Novakovich},\ and\ \citenamefont
  {Kochetov}}]{TNK+18}%
  \BibitemOpen
  \bibfield  {author} {\bibinfo {author} {\bibfnamefont {A.}~\bibnamefont
  {Taranukhina}}, \bibinfo {author} {\bibfnamefont {A.}~\bibnamefont
  {Novakovich}}, \ and\ \bibinfo {author} {\bibfnamefont {V.}~\bibnamefont
  {Kochetov}},\ }in\ \href@noop {} {\emph {\bibinfo {booktitle} {Multiple
  Scattering Theory for Spectroscopies: A Guide to Multiple Scattering Computer
  Codes}}},\ \bibinfo {editor} {edited by\ \bibinfo {editor} {\bibfnamefont
  {D.}~\bibnamefont {S\'{e}billeau}}, \bibinfo {editor} {\bibfnamefont
  {K.}~\bibnamefont {Hatada}}, \ and\ \bibinfo {editor} {\bibfnamefont
  {H.}~\bibnamefont {Ebert}}}\ (\bibinfo  {publisher} {Springer},\ \bibinfo
  {address} {Berlin},\ \bibinfo {year} {2018})\ Chap.~\bibinfo {chapter} {13},
  pp.\ \bibinfo {pages} {309--315}\BibitemShut {NoStop}%
\bibitem [{\citenamefont {Joly}(2015)}]{fdmnes-code}%
  \BibitemOpen
  \bibfield  {author} {\bibinfo {author} {\bibfnamefont {Y.}~\bibnamefont
  {Joly}},\ }\href@noop {} {\emph {\bibinfo {title} {The {\sc fnmnes} code}}},\
  \bibinfo {address} {\url{http://neel.cnrs.fr/spip.php?rubrique1007&lang=en}}
  (\bibinfo {year} {2015})\BibitemShut {NoStop}%
\bibitem [{\citenamefont {Bun\u{a}u}\ and\ \citenamefont {Joly}(2009)}]{BJ+09}%
  \BibitemOpen
  \bibfield  {author} {\bibinfo {author} {\bibfnamefont {O.}~\bibnamefont
  {Bun\u{a}u}}\ and\ \bibinfo {author} {\bibfnamefont {Y.}~\bibnamefont
  {Joly}},\ }\href {http://stacks.iop.org/0953-8984/21/i=34/a=345501}
  {\bibfield  {journal} {\bibinfo  {journal} {J. Phys.: Condens. Matter}\
  }\textbf {\bibinfo {volume} {21}},\ \bibinfo {pages} {345501} (\bibinfo
  {year} {2009})}\BibitemShut {NoStop}%
\bibitem [{\citenamefont {Rehr}(2013)}]{feff-code}%
  \BibitemOpen
  \bibfield  {author} {\bibinfo {author} {\bibfnamefont {J.~J.}\ \bibnamefont
  {Rehr}},\ }\href@noop {} {\emph {\bibinfo {title} {The {\sc feff} code,
  version 9}}},\ \bibinfo {address} {\url{http://feffproject.org}} (\bibinfo
  {year} {2013})\BibitemShut {NoStop}%
\bibitem [{\citenamefont {Rehr}\ \emph {et~al.}(2009)\citenamefont {Rehr},
  \citenamefont {Kas}, \citenamefont {Prange}, \citenamefont {Sorini},
  \citenamefont {Takimoto},\ and\ \citenamefont {Vila}}]{RKP+09}%
  \BibitemOpen
  \bibfield  {author} {\bibinfo {author} {\bibfnamefont {J.~J.}\ \bibnamefont
  {Rehr}}, \bibinfo {author} {\bibfnamefont {J.~J.}\ \bibnamefont {Kas}},
  \bibinfo {author} {\bibfnamefont {M.~P.}\ \bibnamefont {Prange}}, \bibinfo
  {author} {\bibfnamefont {A.~P.}\ \bibnamefont {Sorini}}, \bibinfo {author}
  {\bibfnamefont {Y.}~\bibnamefont {Takimoto}}, \ and\ \bibinfo {author}
  {\bibfnamefont {F.}~\bibnamefont {Vila}},\ }\href {\doibase
  http://dx.doi.org/10.1016/j.crhy.2008.08.004} {\bibfield  {journal} {\bibinfo
   {journal} {C. R. Phys.}\ }\textbf {\bibinfo {volume} {10}},\ \bibinfo
  {pages} {548} (\bibinfo {year} {2009})}\BibitemShut {NoStop}%
\bibitem [{\citenamefont {S\'{e}billeau}(2017)}]{msspec-code}%
  \BibitemOpen
  \bibfield  {author} {\bibinfo {author} {\bibfnamefont {D.}~\bibnamefont
  {S\'{e}billeau}},\ }\href@noop {} {\emph {\bibinfo {title} {The {\sc MsSpec}
  code}}},\ \bibinfo {address}
  {\url{https://ipr.univ-rennes1.fr/msspec?lang=en}} (\bibinfo {year}
  {2017})\BibitemShut {NoStop}%
\bibitem [{\citenamefont {S\'{e}billeau}\ \emph {et~al.}(2011)\citenamefont
  {S\'{e}billeau}, \citenamefont {Natoli}, \citenamefont {Gavaza},
  \citenamefont {Zhao}, \citenamefont {Pieve},\ and\ \citenamefont
  {Hatada}}]{SNG+11}%
  \BibitemOpen
  \bibfield  {author} {\bibinfo {author} {\bibfnamefont {D.}~\bibnamefont
  {S\'{e}billeau}}, \bibinfo {author} {\bibfnamefont {C.}~\bibnamefont
  {Natoli}}, \bibinfo {author} {\bibfnamefont {G.~M.}\ \bibnamefont {Gavaza}},
  \bibinfo {author} {\bibfnamefont {H.}~\bibnamefont {Zhao}}, \bibinfo {author}
  {\bibfnamefont {F.~D.}\ \bibnamefont {Pieve}}, \ and\ \bibinfo {author}
  {\bibfnamefont {K.}~\bibnamefont {Hatada}},\ }\href {\doibase
  http://dx.doi.org/10.1016/j.cpc.2011.07.012} {\bibfield  {journal} {\bibinfo
  {journal} {Comp. Phys. Commun.}\ }\textbf {\bibinfo {volume} {182}},\
  \bibinfo {pages} {2567} (\bibinfo {year} {2011})}\BibitemShut {NoStop}%
\bibitem [{\citenamefont {Benfatto}\ and\ \citenamefont {{Della
  Longa}}(2003)}]{mxan-code}%
  \BibitemOpen
  \bibfield  {author} {\bibinfo {author} {\bibfnamefont {M.}~\bibnamefont
  {Benfatto}}\ and\ \bibinfo {author} {\bibfnamefont {S.}~\bibnamefont {{Della
  Longa}}},\ }\href@noop {} {\emph {\bibinfo {title} {The {\sc mxan} code}}},\
  \bibinfo {address}
  {\url{http://http://www.esrf.eu/computing/scientific/MXAN}} (\bibinfo {year}
  {2003})\BibitemShut {NoStop}%
\bibitem [{\citenamefont {Benfatto}\ \emph {et~al.}(2003)\citenamefont
  {Benfatto}, \citenamefont {{Della Longa}},\ and\ \citenamefont
  {Natoli}}]{BDN+03}%
  \BibitemOpen
  \bibfield  {author} {\bibinfo {author} {\bibfnamefont {M.}~\bibnamefont
  {Benfatto}}, \bibinfo {author} {\bibfnamefont {S.}~\bibnamefont {{Della
  Longa}}}, \ and\ \bibinfo {author} {\bibfnamefont {C.~R.}\ \bibnamefont
  {Natoli}},\ }\href@noop {} {\bibfield  {journal} {\bibinfo  {journal} {J.\
  Synchr.\ Rad.}\ }\textbf {\bibinfo {volume} {10}},\ \bibinfo {pages} {51}
  (\bibinfo {year} {2003})}\BibitemShut {NoStop}%
\bibitem [{\citenamefont {Ebert}(2017)}]{sprkkr-code}%
  \BibitemOpen
  \bibfield  {author} {\bibinfo {author} {\bibfnamefont {H.}~\bibnamefont
  {Ebert}},\ }\href@noop {} {\emph {\bibinfo {title} {The {\sc sprkkr} code,
  version 7.7}}},\ \bibinfo {address}
  {\url{http://ebert.cup.uni-muenchen.de/SPRKKR}} (\bibinfo {year}
  {2017})\BibitemShut {NoStop}%
\bibitem [{\citenamefont {Ebert}\ \emph {et~al.}(2011)\citenamefont {Ebert},
  \citenamefont {K\"odderitzsch},\ and\ \citenamefont {Min\'{a}r}}]{EKM11}%
  \BibitemOpen
  \bibfield  {author} {\bibinfo {author} {\bibfnamefont {H.}~\bibnamefont
  {Ebert}}, \bibinfo {author} {\bibfnamefont {D.}~\bibnamefont
  {K\"odderitzsch}}, \ and\ \bibinfo {author} {\bibfnamefont {J.}~\bibnamefont
  {Min\'{a}r}},\ }\href@noop {} {\bibfield  {journal} {\bibinfo  {journal}
  {Rep. Prog. Phys.}\ }\textbf {\bibinfo {volume} {74}},\ \bibinfo {pages}
  {096501} (\bibinfo {year} {2011})}\BibitemShut {NoStop}%
\bibitem [{\citenamefont {Zeller}(1988)}]{Zel88}%
  \BibitemOpen
  \bibfield  {author} {\bibinfo {author} {\bibfnamefont {R.}~\bibnamefont
  {Zeller}},\ }\href {\doibase 10.1007/BF01313114} {\bibfield  {journal}
  {\bibinfo  {journal} {Z. Physik B}\ }\textbf {\bibinfo {volume} {72}},\
  \bibinfo {pages} {79} (\bibinfo {year} {1988})}\BibitemShut {NoStop}%
\bibitem [{\citenamefont {\v{S}ipr}\ \emph {et~al.}(2011)\citenamefont
  {\v{S}ipr}, \citenamefont {Min\'ar}, \citenamefont {Scherz}, \citenamefont
  {Wende},\ and\ \citenamefont {Ebert}}]{SMS+11}%
  \BibitemOpen
  \bibfield  {author} {\bibinfo {author} {\bibfnamefont {O.}~\bibnamefont
  {\v{S}ipr}}, \bibinfo {author} {\bibfnamefont {J.}~\bibnamefont {Min\'ar}},
  \bibinfo {author} {\bibfnamefont {A.}~\bibnamefont {Scherz}}, \bibinfo
  {author} {\bibfnamefont {H.}~\bibnamefont {Wende}}, \ and\ \bibinfo {author}
  {\bibfnamefont {H.}~\bibnamefont {Ebert}},\ }\href {\doibase
  10.1103/PhysRevB.84.115102} {\bibfield  {journal} {\bibinfo  {journal} {Phys.
  Rev. B}\ }\textbf {\bibinfo {volume} {84}},\ \bibinfo {pages} {115102}
  (\bibinfo {year} {2011})}\BibitemShut {NoStop}%
\bibitem [{\citenamefont {Campbell}\ and\ \citenamefont {Papp}(2001)}]{CP01}%
  \BibitemOpen
  \bibfield  {author} {\bibinfo {author} {\bibfnamefont {J.~L.}\ \bibnamefont
  {Campbell}}\ and\ \bibinfo {author} {\bibfnamefont {T.}~\bibnamefont
  {Papp}},\ }\href@noop {} {\bibfield  {journal} {\bibinfo  {journal} {At. Data
  Nucl. Data Tables}\ }\textbf {\bibinfo {volume} {7}},\ \bibinfo {pages} {1}
  (\bibinfo {year} {2001})}\BibitemShut {NoStop}%
\bibitem [{\citenamefont {\v{S}ipr}\ and\ \citenamefont {Ebert}(2005)}]{SE05}%
  \BibitemOpen
  \bibfield  {author} {\bibinfo {author} {\bibfnamefont {O.}~\bibnamefont
  {\v{S}ipr}}\ and\ \bibinfo {author} {\bibfnamefont {H.}~\bibnamefont
  {Ebert}},\ }\href@noop {} {\bibfield  {journal} {\bibinfo  {journal} {Phys.
  Rev. B}\ }\textbf {\bibinfo {volume} {72}},\ \bibinfo {pages} {134406}
  (\bibinfo {year} {2005})}\BibitemShut {NoStop}%
\bibitem [{\citenamefont {M\"uller}\ \emph {et~al.}(1982)\citenamefont
  {M\"uller}, \citenamefont {Jepsen},\ and\ \citenamefont {Wilkins}}]{MJW82}%
  \BibitemOpen
  \bibfield  {author} {\bibinfo {author} {\bibfnamefont {J.~E.}\ \bibnamefont
  {M\"uller}}, \bibinfo {author} {\bibfnamefont {O.}~\bibnamefont {Jepsen}}, \
  and\ \bibinfo {author} {\bibfnamefont {J.~W.}\ \bibnamefont {Wilkins}},\
  }\href@noop {} {\bibfield  {journal} {\bibinfo  {journal} {Solid State
  Commun.}\ }\textbf {\bibinfo {volume} {42}},\ \bibinfo {pages} {365}
  (\bibinfo {year} {1982})}\BibitemShut {NoStop}%
\bibitem [{\citenamefont {Beni}\ and\ \citenamefont {Platzman}(1976)}]{BP+76}%
  \BibitemOpen
  \bibfield  {author} {\bibinfo {author} {\bibfnamefont {G.}~\bibnamefont
  {Beni}}\ and\ \bibinfo {author} {\bibfnamefont {P.~M.}\ \bibnamefont
  {Platzman}},\ }\href@noop {} {\bibfield  {journal} {\bibinfo  {journal}
  {Phys. Rev. B}\ }\textbf {\bibinfo {volume} {14}},\ \bibinfo {pages} {1514}
  (\bibinfo {year} {1976})}\BibitemShut {NoStop}%
\bibitem [{\citenamefont {Fujikawa}\ \emph {et~al.}(1999)\citenamefont
  {Fujikawa}, \citenamefont {Rehr}, \citenamefont {Wada},\ and\ \citenamefont
  {Nagamatsu}}]{FRW+99}%
  \BibitemOpen
  \bibfield  {author} {\bibinfo {author} {\bibfnamefont {T.}~\bibnamefont
  {Fujikawa}}, \bibinfo {author} {\bibfnamefont {J.}~\bibnamefont {Rehr}},
  \bibinfo {author} {\bibfnamefont {Y.}~\bibnamefont {Wada}}, \ and\ \bibinfo
  {author} {\bibfnamefont {S.}~\bibnamefont {Nagamatsu}},\ }\href {\doibase
  10.1143/JPSJ.68.1259} {\bibfield  {journal} {\bibinfo  {journal} {J. Phys.
  Soc. Japan}\ }\textbf {\bibinfo {volume} {68}},\ \bibinfo {pages} {1259}
  (\bibinfo {year} {1999})}\BibitemShut {NoStop}%
\bibitem [{\citenamefont {{\v{S}}ipr}\ \emph {et~al.}(2016)\citenamefont
  {{\v{S}}ipr}, \citenamefont {Vack{\'{a}}{\v{r}}},\ and\ \citenamefont
  {Kuzmin}}]{SVK+16}%
  \BibitemOpen
  \bibfield  {author} {\bibinfo {author} {\bibfnamefont {O.}~\bibnamefont
  {{\v{S}}ipr}}, \bibinfo {author} {\bibfnamefont {J.}~\bibnamefont
  {Vack{\'{a}}{\v{r}}}}, \ and\ \bibinfo {author} {\bibfnamefont
  {A.}~\bibnamefont {Kuzmin}},\ }\href {\doibase 10.1107/S1600577516014570}
  {\bibfield  {journal} {\bibinfo  {journal} {J.\ Synchr.\ Rad.}\ }\textbf
  {\bibinfo {volume} {23}},\ \bibinfo {pages} {1433} (\bibinfo {year}
  {2016})}\BibitemShut {NoStop}%
\bibitem [{\citenamefont {Shamma}\ \emph {et~al.}(1992)\citenamefont {Shamma},
  \citenamefont {Abbate},\ and\ \citenamefont {Fuggle}}]{SAB+92}%
  \BibitemOpen
  \bibfield  {author} {\bibinfo {author} {\bibfnamefont {F.~A.}\ \bibnamefont
  {Shamma}}, \bibinfo {author} {\bibfnamefont {M.}~\bibnamefont {Abbate}}, \
  and\ \bibinfo {author} {\bibfnamefont {J.~C.}\ \bibnamefont {Fuggle}},\ }in\
  \href@noop {} {\emph {\bibinfo {booktitle} {Unoccupied Electron States}}},\
  \bibinfo {editor} {edited by\ \bibinfo {editor} {\bibfnamefont {J.~C.}\
  \bibnamefont {Fuggle}}\ and\ \bibinfo {editor} {\bibfnamefont {J.~E.}\
  \bibnamefont {Inglesfield}}}\ (\bibinfo  {publisher} {Springer Verlag},\
  \bibinfo {address} {Berlin},\ \bibinfo {year} {1992})\ p.\ \bibinfo {pages}
  {347}\BibitemShut {NoStop}%
\bibitem [{\citenamefont {Lee}\ \emph {et~al.}(1981)\citenamefont {Lee},
  \citenamefont {Citrin}, \citenamefont {Eisenberger},\ and\ \citenamefont
  {Kincaid}}]{LCE+81}%
  \BibitemOpen
  \bibfield  {author} {\bibinfo {author} {\bibfnamefont {P.~A.}\ \bibnamefont
  {Lee}}, \bibinfo {author} {\bibfnamefont {P.~H.}\ \bibnamefont {Citrin}},
  \bibinfo {author} {\bibfnamefont {P.}~\bibnamefont {Eisenberger}}, \ and\
  \bibinfo {author} {\bibfnamefont {B.~M.}\ \bibnamefont {Kincaid}},\ }\href
  {\doibase 10.1103/RevModPhys.53.769} {\bibfield  {journal} {\bibinfo
  {journal} {Rev. Mod. Phys.}\ }\textbf {\bibinfo {volume} {53}},\ \bibinfo
  {pages} {769} (\bibinfo {year} {1981})}\BibitemShut {NoStop}%
\end{thebibliography}%

%%%\bibliography{liter-broadening}

% File *.bbl inserted manually in order to avoid need for BibTeX
% cooperation

\end{document}